\documentclass[12pt,a4paper]{article}
\usepackage{amsmath}
\usepackage{amssymb}
\usepackage{graphicx}
\begin{document}
 \begin{titlepage}
  \begin{flushright}
IFUP--TH/2011-7-r\\
  \end{flushright}
~
\vskip .8truecm
 \begin{center}
\Large\bf
Riemann-Hilbert treatment of Liouville theory on the torus: The general case
\end{center}
\vskip 1.2truecm
\begin{center}
{Pietro Menotti} \\
{\small\it Dipartimento di Fisica, Universit{\`a} di Pisa and}\\
{\small\it INFN, Sezione di Pisa, Largo B. Pontecorvo 3, I-56127}\\
{\small\it e-mail: menotti@df.unipi.it}\\
 \end{center}
\vskip 1.2truecm
\centerline{April 2011}
                
\vskip 1.2truecm
                                                              
\begin{abstract}
We extend the previous treatment of Liouville theory on the torus, to the
general case in which the distribution of charges is not necessarily
symmetric. This requires the concept of Fuchsian differential equation on
Riemann surfaces. We show through a group theoretic argument that the Heun
parameter and a weight constant are sufficient to satisfy all monodromy
conditions. We then apply the technique of differential equation on a Riemann
surface to the two point function on the torus in which one source is
arbitrary and the other small. As a byproduct we give in terms of quadratures
the exact Green function on the square and on the rhombus with opening angle
$2\pi/6$ in the background of the field generated by an arbitrary charge.
\end{abstract}

\end{titlepage}

\eject

\section{Introduction}

Liouville theory plays a pivotal role in several fields both at the classical
\cite{picard,poincare,lichtenstein,troyanov} and
quantum level \cite{thorn,dorn,teschner,ZZ,ZZdisk,weigt,nakayama}. 
Liouville action provides the Faddeev-Popov determinant 
in two dimensional gravity \cite{polyakov} and it
gives the complete hamiltonian structure of 
$2+1$-dimensional gravity coupled with particles \cite{cantini}. 
It appears also in the soliton solutions to the gauged non linear Schroedinger
equation on the plane
\cite{olesen,jackiw,akerblom}. Liouville field
has been suggested to play the role of a fifth dimension in the AdS-CFT
treatment of QCD \cite{gubser}.

Recently a renewed interest has developed due to a conjecture \cite{gaiotto}
that Liouville theory on a Riemann surface of genus $g$ is related to a
certain class of ${\cal N}=2$, $4$-dimensional gauge theories and the
conjecture has been supported by extensive tests on genus $0$ and $1$
\cite{alday,drukker,alba} and proven in a class of cases \cite{hadasz}.

In the topology of the sphere important results regarding the four point
conformal correlation functions have been obtained in 
\cite{onofri}
and their relation to the one point function on the torus suggested 
in \cite{onofri} and proven in \cite{hadasz2}.

We recall that the classical solution provides the starting
point for the semiclassical expansion and that such semiclassical expansion 
was used to confirm the first few terms of the bootstrap solution on the
sphere, pseudosphere and disk topologies and also to give some results on 
higher
point functions when one source is weak \cite{MV,MT,menotti}.
 
The simplest situation is given by the topology of the sphere for which many
results are available both at the classical and quantum level. The torus
topology is intrinsically more complicated that the sphere. For example
the one point function on the torus appears to be of comparable complexity
as the four point function on the sphere \cite{onofri}. 

In a previous paper \cite{menottitorus} Liouville theory on the torus was 
examined in
the simpler situation in which the distribution of charges is invariant under
reflection $z\rightarrow -z$. As the map given by the Weierstrass function
$u=\wp(z)$ is invariant under reflection it turns out that one can map the
problem on the Riemann sphere, where well developed techniques exist along
with several results. In the symmetric situation, periodicity, which is the
fundamental constraint, is obtained by imposing monodromicity in $u$. In fact
monodromic behavior in $u$ at the singular points points $e_k$ (see
\cite{menottitorus} and section \ref{thetorus} of the present paper) 
combined with reflection symmetry is equivalent to the
periodicity constraint along the non collapsible cycles which in the 
general case is much more difficult requirement to implement.

In \cite{menottitorus} the exact solution for the one point function on the
square 
was given and several problems which can be dealt with perturbation technique
solved.

In the present paper we deal with the problem in full generality i.e. when the
distribution of charges is not necessarily invariant under reflections; to do
this the shall need the full description of the torus as a double sheet
cut-plane and we shall need the notion \cite{katz,gallo,hawley} of 
differential equation on
a Riemann surface. As discussed in \cite{menottitorus} the differential 
equation
which solves the Liouville equation in the case of the torus contains even in
the simplest case of the one point function a parameter, the Heun parameter
which does not appear in the case of the sphere topology with three
sources. Such parameter, along with an other weight parameter has to be
determined by imposing monodromicity on the two sheet cut plane or
equivalently by imposing the periodic boundary conditions. In the case of the
square and of the rhombus with opening angle $2\pi/6$ such parameter turns
out to be zero and the problem is reduced to an hypergeometric equation (see
Appendix B); on the other hand the exact non perturbative determination of
the Heun parameter in the general case to our knowledge in not an accomplished
task \cite{erdelyi, hejhal, hempel,smith}.

The structure of the paper is the following:
In section \ref{general} we display the mathematical framework of the 
formulation of differential equations on a Riemann surface.

In section \ref{thetorus} we show with a group theory argument how the Heun 
parameter along
with the weight parameter $|\kappa|$ give the necessary and sufficient degrees 
of freedom to satisfy all the monodromy requirements. 

Then in section \ref{additionweak} we solve the problem of the addition of a 
small charge when
the solution for a single (non necessarily small) charge is given, as it is
the case of the square and of the rhombus with opening angle $2\pi/6$. The
treatment however is completely general.

As a by product we obtain is section \ref{greenfunction} the expression by
quadratures of the exact Green function on the background generated by an
arbitrary charge. The expression holds also for the general torus; the only
difference is that while for the square and the rhombus of opening angle
$2\pi/6$ we know the expression of the unperturbed solution, for
the generic torus such unperturbed solutions are not known in terms of usual
functions.

In order to display the workings of the technique discussed in sections
\ref{general} and \ref{thetorus}, in Appendix A we give the perturbative
determination of the conformal factor for a weak source in the general case
i.e. when the source is not symmetric wrt to the position of the kinematical
singularities $\omega_k$ which arise in the transition from the global
covering variable to the coordinates $(u=\wp(z),w=\wp'(z))$ which provide a
one-to-one description of the torus.
  
In Appendix B for completeness we report the solution for the square
given in \cite{menottitorus} and we also add the solution in terms of
hypergeometric functions for the rhombus of opening angle $2\pi/6$. The
deformation technique developed in \cite{menottitorus} can be applied to both
of these cases.

\bigskip

\section{Differential equations on a Riemann surface}\label{general}

We give here the elements of the theory of differential equations on a Riemann
surface. For more details see e.g. \cite{katz,gallo,hawley}. 
A surface with the topology of
the sphere can be mapped on the compactified plane where the usual theory of
linear differential equations apply (see e.g. 
\cite{poole,hille,okamoto,yoshida,jimbo}).
To extend the concept of differential equation to a Riemann surface it is 
useful to rewrite the second
order differential equation 
\begin{equation}\label{secondorderDE}
y''(x) + q(x)y(x) =0
\end{equation}
in the form
\begin{equation}
dY + \Gamma Y = \frac{dY}{dx} dx+ \Gamma_x ~dx~ Y =0
\end{equation}
where
\begin{equation}
Y =
\begin{pmatrix}
y(x)\\y^{(1)}(x)
\end{pmatrix}
\end{equation}
and
\begin{equation}
\Gamma_x=
\begin{pmatrix}
0&-1\\q(x) &0
\end{pmatrix}
\end{equation}

and we have denoted in eq.(\ref{secondorderDE}) by a prime the differentiation 
wrt $x$. For well known
topological reasons when the manifold has genus $1$ or higher we have no
global coordinate at our disposal and thus we must use more than one chart
with analytic transition functions. 

It is useful to consider $y$ as a differential of the fractionary order 
$-\frac{1}{2}$, i.e. in the change from the variable $x$ to the variable
$\tilde x$ we have
\begin{equation}\label{minusonehalfdiff}
\tilde y(\tilde x)=
y(x(\tilde x))\left(\frac{dx}{d\tilde x}\right)
^{-\frac{1}{2}}\equiv y(x(\tilde x)) s(\tilde x)
\end{equation}
where
\begin{equation}
s(\tilde x)\equiv \left(\frac{dx}{d\tilde x}\right)
^{-\frac{1}{2}}.
\end{equation}
One easily finds, denoting with a dot the derivative wrt $\tilde x$
\begin{equation}
\tilde Y(\tilde x) =
\begin{pmatrix}
s(\tilde x) & 0\\ \dot s(\tilde x) &\frac{1}{s(\tilde x)}
\end{pmatrix}
Y(x) \equiv U^{-1}Y(x),~~~~U\in SL(2 C)
\end{equation}
and $\tilde Y$ obeys
\begin{equation}
d\tilde Y + \tilde \Gamma \tilde Y = \frac{d\tilde Y}{d\tilde x} d\tilde
x+ \tilde \Gamma_{\tilde x} ~d\tilde x~ \tilde Y =0.
\end{equation}
The usual transformation of a connection gives
\begin{equation}
\tilde \Gamma = U^{-1} dU +U^{-1} \Gamma U = \tilde\Gamma_{\tilde x}~d\tilde x
\end{equation}
with
\begin{equation}\label{gammatransform}
\tilde \Gamma_{\tilde x } =
\begin{pmatrix}
0&-1\\
\frac{q}{s^4}-\frac{\ddot s}{s}& 0
\end{pmatrix}.
\end{equation}
The term
$\frac{\ddot s}{s}$ is the Schwarzian derivative of the transformation
\begin{equation}
\frac{\ddot s}{s}=\{x,\tilde x\}=
\frac{3}{4}\left(\frac{\ddot x}{\dot x}\right)^2
-\frac{1}{2}\frac{\dddot x}{\dot x}=-\{\tilde x, x\}\left(\frac{d\tilde x}{dx}
\right)^{-2}
\end{equation}
It is of interest to notice that the assumed nature (\ref{minusonehalfdiff}) 
of the solutions 
$y$ of eq.(\ref{secondorderDE}), which leaves the Wronskian of two solutions
unchanged, 
makes the well known expression \cite{menottitorus} for the conformal factor
in terms of two 
independent solutions of the differential equation (\ref{secondorderDE})
\begin{equation}
e^{\phi}=\frac{2 |w_{12}|^2}{[\overline{y_1(x)}y_1(x)
-\overline{y_2(x)} y_2(x)]^2}
 \end{equation}
a differential of order $(1,1)$ as required for an area element
\begin{equation}
e^{\phi(x)} dx\wedge d\bar x=
e^{\tilde\phi(\tilde x)} d\tilde x\wedge d\bar {\tilde x}~.
\end{equation} 
One can extend the concept of Fuchsian differential equation to differential
equations on a
Riemann surface provided the term $q$ or better the $(2,1)$ component of the
connection $\Gamma$ is a meromorphic function on the Riemann surface
i.e. \cite{GH} the ratio of two polynomials of the same order in the
homogeneous coordinates describing the surface.

\bigskip

\section{The torus}\label{thetorus}

The torus is described by the cubic curve in euclidean coordinates \cite{GH} 
\begin{equation}\label{cubic}
(u,w),~~~~~ w^2 = 4(u-e_1)(u-e_2)(u-e_3)=4 u^3 -g_2 u - g_3
\end{equation}
with
\begin{equation}
e_1+e_2+e_3=0
\end{equation}
or in homogeneous coordinates
\begin{equation}
X_2^2 X_0= 4(X_1-e_1 X_0)(X_1-e_2X_0)(X_1-e_3X_0).
\end{equation}

To the Liouville equation
\begin{equation}\label{liouvilleequation}
-\partial_z\partial_{\bar z}\phi+e^{\phi}= 2\pi \eta\delta(z-z_t)
\end{equation}
there corresponds the differential equation in $u$ given by
\begin{equation}
y''(u)+ Q(u,w) y(u)=0
\end{equation}
with $Q(u,w)$ a meromorphic function on the torus with second order 
poles at $e_k$ due to the presence of the Schwarz derivative in 
(\ref{gammatransform}) and a
second order pole at $(u,w)=(t,w_t)$
\begin{eqnarray}\label{Q}
Q(u,w)&=&\frac{3}{16}\left(\frac{1}{(u-e_1)^2}+ \frac{1}{(u-e_2)^2}+
\frac{1}{(u-e_3)^2}+\frac{2 e_1}{(e_1-e_2)(e_3-e_1)(u-e_1)}\right.\nonumber\\
&+& \left.\frac{2 e_2}{(e_2-e_3)(e_1-e_2)(u-e_2)} 
+\frac{2 e_3}{(e_3-e_1)(e_2-e_3)(u-e_3)}\right) \nonumber\\
&+&\frac{1-\lambda^2}{4 w_t}
\left[\frac{(w+w_t)^2}{4(u-t)^2}
-u-2t\right]\frac{1}{w} 
+\frac{\beta_t(w+w_t)}{2(u-t)w}\nonumber\\
&+&\frac{\beta_1}{2(u-e_1)}+\frac{\beta_2}{2(u-e_2)}+\frac{\beta_3}{2(u-e_3)}
\end{eqnarray}
where the $\lambda$ appearing in the residue of the double pole at 
$(u,w)=(t,w_t)$ 
is given by $\lambda =1-2\eta$. Due to the factor $w+w_t$ the pole is present  
only on the first sheet. The term $-u-2t$ is necessary to assure that $Q$ does
not show an irregular singularity at infinity. 
The $\beta_1,\beta_2,\beta_3, \beta_t$ are the accessory parameters. 
The above structure can be trivially generalized to any finite number of 
sources. Obviously we could place the source at $z_t=0$ simplifying the
structure of $Q$ as was done in 
\cite{menottitorus} but in the perspective of the general non symmetric
situation (see e.g. section \ref{additionweak}) we work here in full
generality. 

The first two lines 
in the expression of $Q$ is simply the contribution of the Schwarzian
derivative for the transition from the covering variable $z$ to the variable
$u$.  
The asymptotic behavior of $Q$ in the local covering variable at infinity $v$, 
with
$v^2=1/u$ is given by
\begin{eqnarray}
& &\frac{1}{2}(\beta_1+\beta_2+\beta_3+\beta_t)v^2 +
v^4\left(\frac{3}{16}+\frac{1}{2}\left(\beta_1e_1+\beta_2e_2+\beta_3e_3+
\beta_t t +\frac{1-\lambda^2}{4}\right)\right)\nonumber\\
& & + v^5\left(\frac{\beta_t w_t}{4}+
\frac{1-\lambda^2}{32 w_t}(12 t^2- g_2)\right)+O(v^6). 
\end{eqnarray}
Regularity of
\begin{equation}
Q_v(v) =Q(u,w) \left(\frac{du}{dv}\right)^2 -\{u,v\}
\end{equation}
at $v=0$ i.e. absence of charges at $z=0$ implies
\begin{equation}
\beta_1+\beta_2+\beta_3+\beta_t=0
\end{equation}
\begin{equation}
\beta_1e_1+\beta_2e_2+\beta_3e_3+\beta_t t + \frac{1-\lambda^2}{4}=0
\end{equation}
\begin{equation}\label{thirdfuchs}
\frac{\beta_t w_t}{4}+\frac{1-\lambda^2}{32 w_t}(12 t^2- g_2)=0
\end{equation}
and thus at the end we have one free accessory parameter, say $\beta_1$ which
has to be used to satisfy the monodromies at $(t,w_t)$ and along the two 
fundamental non contractible cycles. 
We want to show at the non perturbative level how one free accessory
parameter is sufficient for the purpose.

Let $y_1$ and $y_2$ be two independent solutions of
\begin{equation}
y''+Q(u,w)y=0
\end{equation}
of Wronskian $w_{12}$. We must find combinations
\begin{equation}
y^c_1=a_{11}y_1+a_{12}y_2,~~~~y^c_2=a_{21}y_1+a_{22}y_2,
\end{equation}
such that
\begin{equation}\label{standardform}
|y^c_1|^2-|y^c_2|^2
\end{equation}
is monodromic. As the Wronskian is fixed the transformation must be of
$SL(2,C)$ type, which corresponds to six real degrees of freedom. In
addition we notice that the monodromy of (\ref{standardform}) is unchanged
under an $SU(1,1)$ transformation which corresponds to three real degrees of
freedom. Thus the $a_{jk}$ provide effectively only three real degrees of
freedom to which we have to add the real and imaginary part of $\beta_1$,
i.e. $5$ real degrees of freedom.
We shall have to satisfy the monodromy condition  at all singularities
$(t,w_t)$, $(e_k,0)$ and along the two fundamental non contractible cycles
$C_1$ and $C_2$. 

With regard to the cycle $C_1$
we can use three degrees of freedom (e.g. those given by the $a_{jk}$) 
to obtain  
\begin{equation}\label{MC1}
M_{12}(C_1)=\overline{M}_{21}(C_1)~~~~{\rm and}~~~~
|M_{11}(C_1)|=|M_{22}(C_1)|
\end{equation}
which is sufficient to give the $SU(1,1)$ nature of the transformation
$M(C_1)$. Finally
we use the remaining two real degrees of freedom (the real and imaginary
part of $\beta_1$) to impose
\begin{equation}\label{offdiagonal2}
M_{12}(C_2)=\overline{M}_{21}(C_2)~.
\end{equation}
\begin{figure}[htb]
\begin{center}
\includegraphics{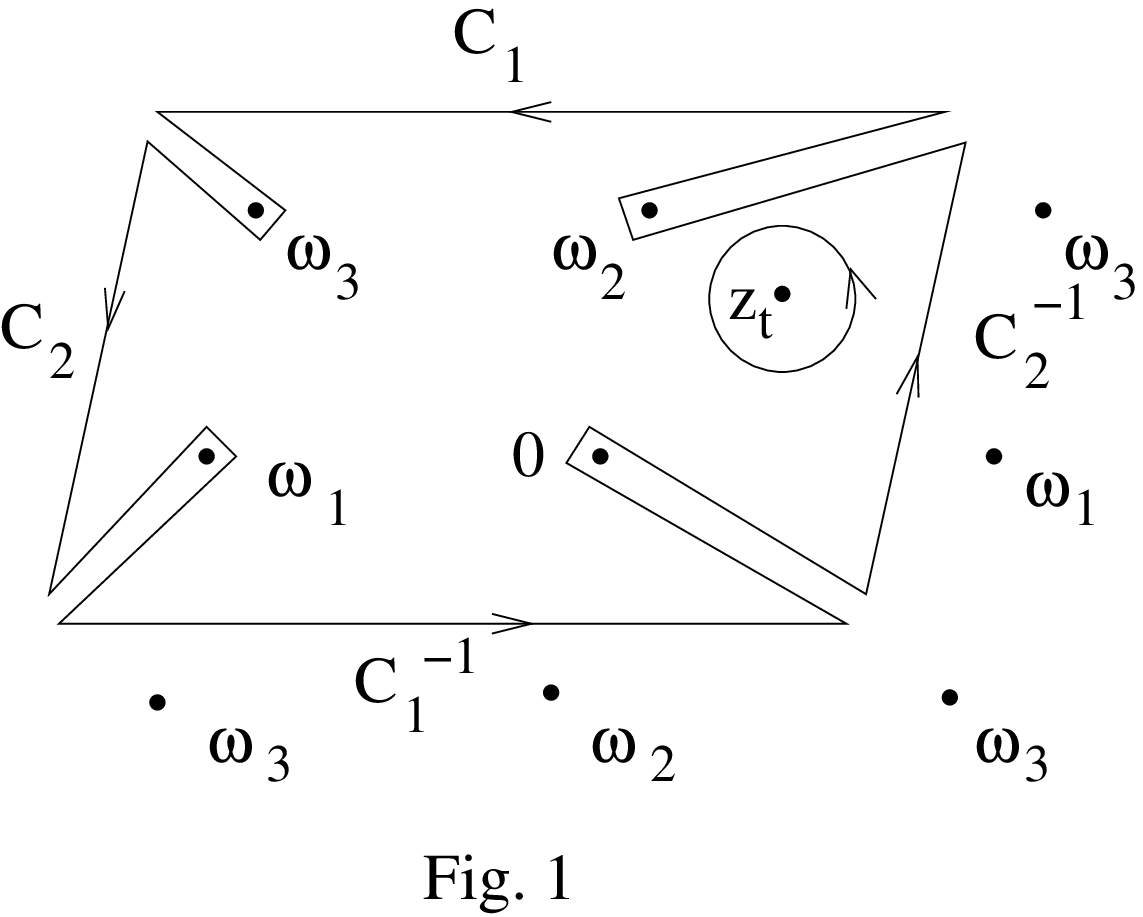}
\end{center}
\end{figure}
This is not enough to give the $SU(1,1)$ nature of the transformation 
$M(C_2)$. However
from the contour shown in Fig.1, obtained by deforming a loop around $z_t$, 
the following relation holds
\begin{equation}\label{nonpertcycle}
M(C_1)M(C_2)M(C_1)^{-1}M(C_2)^{-1} =E_t
\end{equation}
being $E_t$ the elliptic transformation around $z_t$. The reason is that the
monodromy transformations around each of the singular points $z=\omega_k$ 
and $z=0$ is simply $-1$, because the indices at those 
points are $1/4,3/4$ in $u-e_k$ and $-3/4, -1/4$ at $u=\infty$ 
and a turn in the local covering variable corresponds to a double turn in 
$u-e_k$ and in  $u$ at infinity.
Now the following
simple theorem holds: If in eq.(\ref{nonpertcycle}) $M(C_1)\in
SU(1,1)$ and eq.(\ref{offdiagonal2}) is satisfied and $E_t$ is elliptic, 
then $M(C_2)\in SU(1,1)$ and as a consequence also $E_t\in SU(1,1)$. 

In fact after eqs.(\ref{MC1},\ref{offdiagonal2}) 
are satisfied the matrices $M(C_1)$ and $M(C_2)$
take the form
\begin{equation}
M(C_1)=
\begin{pmatrix}
m_{11} & m_{12}\\
\bar m_{12} & \bar m_{11}
\end{pmatrix}
\end{equation}
\begin{equation}
M(C_2)=
\begin{pmatrix}
p e^{i\alpha}/\rho& n_{12}\\
\bar n_{12} & pe^{-i\alpha}\rho
\end{pmatrix}
=
\begin{pmatrix}
n_{11}/\rho& n_{12}\\
\bar n_{12} & \bar n_{11}\rho
\end{pmatrix}.
\end{equation}

The positive number $\rho$ is fixed by the elliptic nature of the product
(\ref{nonpertcycle}) i.e.
\begin{equation}\label{traceconstraint}
{\rm tr}(M(C_1) M(C_2) M(C_1)^{-1}M(C_2)^{-1})= R+ 
A\rho+B \rho^2 + \bar A \rho^{-1}+ \bar B \rho^{-2}={\rm real}
\end{equation}
with
\begin{eqnarray}
R&=& (m_{12}\bar n_{12})^2+m_{11}\bar m_{11}n_{11}\bar n_{11}-
m_{11}^2 n_{12}\bar n_{12} + c.c.={\rm real}\\
A &=&
(\bar m_{11}- m_{11}) \bar n_{11}(\bar n_{12} m_{12}+n_{12} \bar m_{12}),~~~~
B= -m_{12} \bar m_{12} \bar n_{11}^2~.\nonumber
\end{eqnarray}
Eq.(\ref{traceconstraint}) has the discrete solution $\rho=1$. In general
$\rho =1$ is not the only solution of eq.(\ref{traceconstraint}). The
results of Picard \cite{picard,lichtenstein} which apply also to topologies
other than the sphere, 
assure us that the values of the free parameters which realizes the
monodromy are unique and thus $\rho=1$ is the only solution which realizes all
the monodromies.
In the next section we shall apply a perturbative version of this theorem to
the addition of weak sources.

\bigskip

\section{Addition of a weak source}\label{additionweak}

In the present section we shall apply perturbation theory to give in terms of
quadratures the two point conformal factor when one source is arbitrary and 
the second
small. A similar problem was solved in \cite{menottitorus} in the simpler 
situation of the perturbation provided by two weak symmetrical sources and
which
could be treated with methods of the sphere topology. Here instead being the
situation non symmetrical we have to exploit in full the two sheet
representation of the torus given by the cubic (\ref{cubic}). We shall keep the
treatment at full generality; we recall that in two instances ( the square,
and the rhombus of opening angle $2\pi/6$ as given in Appendix B) the
exact solution is known and thus we shall for a perturbation of these
situations 
have a solution in terms of quadratures. The same formulas apply starting from
the general non perturbative one point function with the difference that in
this case the unperturbed functions are given by the solution of the Heun 
equation, at a special value of the Heun parameter for which we do
not possess an explicit formula.

The equation to be solved is
\begin{equation}
y''+ (Q +q)y=0
\end{equation}
where $Q$, describing  a source of arbitrary strength at the origin $z=0$ i.e.
$u=\infty$,
is given by \cite{menottitorus}
\begin{eqnarray}\label{Qterm}
Q(u)&=&
\frac{1-\lambda^2}{16}\frac{u+\beta}{(u-e_1)(u-e_2)(u-e_3)}\\
&+&\frac{3}{16}\left(\frac{1}{(u-e_1)^2}+ \frac{1}{(u-e_2)^2}
+\frac{1}{(u-e_3)^2}+\frac{2 e_1}{(e_1-e_2)(e_3-e_1)(u-e_1)}\right.\nonumber\\
&+& \left.\frac{2 e_2}{(e_2-e_3)(e_1-e_2)(u-e_2)} 
+\frac{2 e_3}{(e_3-e_1)(e_2-e_3)(u-e_3)}\right)\nonumber
\end{eqnarray}
with the non perturbative Heun parameter $\beta$ fixed to the value which 
provides the
monodromic one point solution on the torus. We know the exact value 
of such a parameter only
for the special cases of the square and the rhombus with opening angle
$2\pi/6$ where $\beta=0$ for symmetry reasons. 
We shall denote by $y_1(u)$ and $y_2(u)$ the two unperturbed
solutions, with real Wronskian $w_{12}$, which realize the monodromic 
conformal factor
\begin{equation}\label{unperturbedconformal}
e^{\varphi} du\wedge d\bar u=\frac{2|w_{12}|^2 du\wedge d\bar u}
{[\overline{y_1(u)} y_1(u)-\overline{y_2(u)} y_2(u)]^2}~.             
\end{equation}
The perturbation $q$ is given by
\begin{eqnarray}\label{smallq}
q(u,w) &=&
\frac{\varepsilon}{w_t}
\left[\frac{(w+w_t)^2}{4(u-t)^2} -u-2t\right]\frac{1}{w}
+\frac{\beta_t(w+w_t)}{2(u-t)w}+\nonumber\\
&+&\frac{\beta_1}{2(u-e_1)}+\frac{\beta_2}{2(u-e_2)}+\frac{\beta_3}{2(u-e_3)}
\end{eqnarray}
and it describes the additional weak source at $(t,w_t)$. We had to allow in 
(\ref{smallq}) for a new
accessory parameter $\beta_t$ at $u=t$ and the $\beta_1,\beta_2,\beta_3$ are
the $O(\varepsilon)$ changes of the accessory parameters at
$e_1,e_2,e_3$ of eq.(\ref{Qterm}). Again they are subject to two Fuchs 
relations which are imposed
by the condition that the behavior of $Q+q$ at $u=\infty$ remains unchanged,
i.e. the source at the origin $z=0$ ($u=\infty$) remains unchanged. They are
\begin{equation}
\beta_1+\beta_2+\beta_3+\beta_t=0
\end{equation}
\begin{equation}
\beta_1e_1+\beta_2e_2+\beta_3e_3+\beta_t t+\varepsilon=0.
\end{equation}
The analogue of eq.(\ref{thirdfuchs}) here is absent as now the origin is a 
singular regular point. 
Thus we are left with two free accessory parameters, say $\beta_t$ and
$\beta_1$.
The perturbed solutions are given by
\begin{equation}
y^c_1 = y_1+\delta y_1,~~~~y^c_2= y_2+\delta y_2
\end{equation}
where
\begin{eqnarray}\label{perturbedsolution}
\delta y_1 &=& y_1 \frac{I_{12}}{w_{12}}-y_2 \frac{I_{11}}{w_{12}}+
c_{11}y_1 +c_{12}y_2 \nonumber\\
\delta y_2 &=& y_1 \frac{I_{22}}{w_{12}} - y_2 \frac{I_{12}}{w_{12}}+
c_{21}y_1 +c_{22}y_2 
\end{eqnarray}
with
\begin{equation}
I_{jk}=\int_{u_0}^u q(u,w) y_j(u) y_k(u)du
\end{equation}
and the $O(\varepsilon)$ constants $c_{jk}$ correspond to the addition of the
unperturbed solutions and they satisfy $c_{11}+c_{22}=0$ to leave the
Wronskian unchanged.
As we have already discussed in section 
\ref{thetorus} if one factors 
the $SU(1,1)$ transformations the $c_{jk}$ provide only three real degrees 
of freedom. 
We shall first examine the monodromy at the new singularity $(t,w_t)$.
Taking into account that the integrands of $I_{jk}$ 
contain a first and second order pole at $u=t$ 
whose residues are
\begin{equation}
\beta_t y_j(t) y_k(t)+\varepsilon(y_j(t)y_k(t))'
\end{equation}
we have that the change of such integrals under a tour around $(t,w_t)$ is
\begin{equation}
\delta I_{jk} = 2\pi i\left(\beta_t y_j(t) y_k(t)
+\varepsilon(y_j(t)y_k(t))'\right) .
\end{equation}
The monodromy matrix of $y^c_1$ and $y^c_2$ at $(t,w_t)$ is given by
\begin{equation}\label{Mtmatrix}
M(t)=
\begin{pmatrix}
1+\frac{\delta I_{12}(t)}{w_{12}} & 
-\frac{\delta I_{11}(t)}{w_{12}}\\
\frac{\delta I_{22}(t)}{w_{12}} & 
1-\frac{\delta I_{12}(t)}{w_{12}}   
\end{pmatrix}.
\end{equation}
Monodromy at $(t,w_t)$ requires 
\begin{equation}\label{tcondition}
M_{12}(t)=\overline{M}_{21}(t)~~~~{\rm i.e.}~~~~
1=\frac{\beta_t y_1^2(t)+ 
\varepsilon(y_1^2(t))'}{\bar\beta_t \overline{y_2^2(t)}
+ \varepsilon\overline{(y_2^2(t))'}} ~.
\end{equation}
It is worth noticing that the $\beta_t$ is fixed independently of all 
the other parameters. As the transformation at $(t,w_t)$ is elliptic
i.e. $\varepsilon$ real, the
condition (\ref{tcondition}) is sufficient to assure that the transformation
belongs to $SU(1,1)$. On the other hand $M_{11}(t)=\overline{M}_{22}(t)$
can be explicitly proven using the expression 
of $I_{12}$ as follows. Eq.(\ref{tcondition}) tells us that
\begin{equation}
0<\frac{\beta_t y^2_1(t)+ 
\varepsilon(y_1^2(t))'}{\bar\beta_t \overline{y_2^2(t)}
+ \varepsilon\overline{(y_2^2(t))'}}= \frac{(\beta_t y_1(t)y_2(t)+ 
\varepsilon(y_1(t)y_2(t))')^2-\varepsilon^2w_{12}^2}
{|\beta_t y_2^2(t)+ \varepsilon(y_2^2(t))'|^2}
\end{equation}
which, being $\varepsilon$ and $w_{12}$ real gives
\begin{equation}
\beta_t y_1(t) y_2(t) +\varepsilon (y_1(t) y_2(t))'={\rm real} 
\end{equation}
i.e. $M_{11}(t)=\overline{M}_{22}(t)$.
We add now a remark which will be essential in the following
development. If we consider a contour $D$, see Fig.2, which embraces both the
origin and the
weak singularity at $(t,w_t)$ the resulting monodromy is elliptic. More
precisely the product of the perturbation $M^c_0=M_0+\delta M_0$ 
of the original unperturbed elliptic 
transformation $M_0\in SU(1,1)$ at $z=0$ and the $SU(1,1)$ monodromy near the 
identity at $z=z_t$ is elliptic. In fact
\begin{equation}
M^c_0=
\begin{pmatrix}
m_{11}+\delta m_{11}&m_{12}+\delta m_{12}\\
\bar m_{12}+\delta m_{21}& \bar m_{11}+\delta m_{22}
\end{pmatrix}
\end{equation}
is still elliptic because the source at the origin is unchanged
and ellipticity tells us that $\delta m_{11}+\delta m_{22}={\rm real}$ and 
the elliptic $SU(1,1)$ monodromy $M(t)$ near the identity can be written as
\begin{equation}
M(t)=
\begin{pmatrix}
1+ ir& b\\
\bar b& 1-ir
\end{pmatrix}
\end{equation}
with $r={\rm real}$ and $\delta m_{jk}$, $r$ and $b$ all of order 
$O(\varepsilon)$.
Their product $E$ has the diagonal elements
\begin{equation}
E_{11}=m_{11}+ \delta m_{11}+ ir m_{11}+\bar b m_{12},~~~~
E_{22}=\bar m_{11 }+\delta m_{22}-ir\bar m_{11}+b\bar m_{12}
\end{equation}
with trace real and by continuity, of modulus less than 2. Thus 
$E=M^c_0 M(t)$ is elliptic.

We can now deform the contour $D$ which embraces the origin $z=0$ and $z_t$ as
shown in Fig.2.
\begin{figure}[htb]
\begin{center}
\includegraphics{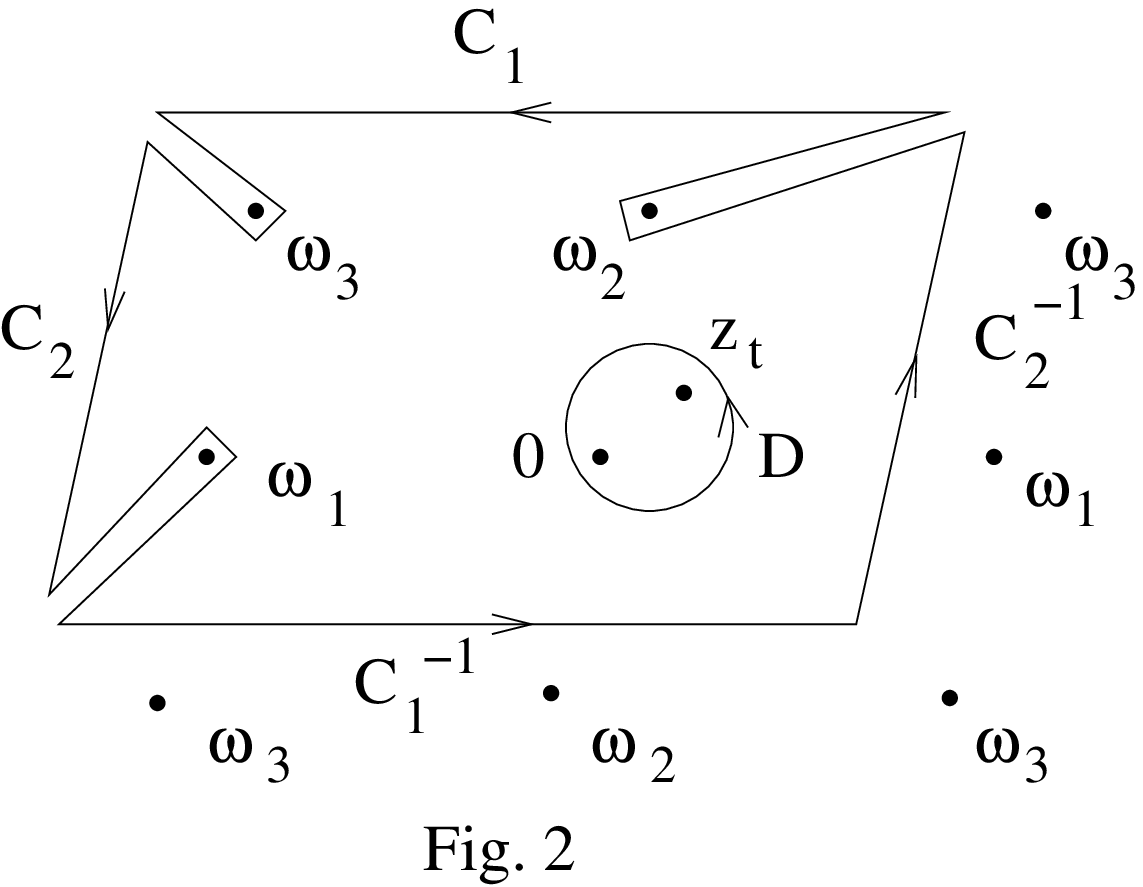}
\end{center}
\end{figure}
Around the singular points $(e_k,0)$ we have
\begin{equation}
\begin{pmatrix}
y_1+\delta y_1 \\
y_2+\delta y_2
\end{pmatrix}\approx
\begin{pmatrix}
1+\frac{I_{12}(e_k)}{w_{12}}+c_{11}&-\frac{I_{11}(e_k)}{w_{12}}+c_{12}\\
\frac{I_{22}(e_k)}{w_{12}}+c_{21}&1-\frac{I_{12}(e_k)}{w_{12}}-c_{11}
\end{pmatrix}
\begin{pmatrix}
y_1\\
y_2
\end{pmatrix}
\end{equation}
due to the cancellation of the $(u-e_k)^{\frac{1}{2}}$ contributions given by
the integrals $I_{jk}$. 
In the local covering variable $w$ we have
\begin{equation}
y_1\approx w^{\frac{1}{2}}(a_1+b_1 w),~~~~
y_2\approx w^{\frac{1}{2}}(a_2+b_2 w)
\end{equation}
so that for a tour around the singularity i.e. $w\rightarrow e^{2 i \pi} w$ 
we have
$y_1\rightarrow -y_1$ and $y_2\rightarrow -y_2$ and thus the monodromy matrices
at $\omega_k$ are all $-1$.  Thus we have
\begin{equation}
M^c(C_1)M^c(C_2)M^c(C_1)^{-1}M^c(C_2)^{-1} = -M^c_0 M(t) = {\rm elliptic}.
\end{equation}
We can now repeat the argument given in
section \ref{thetorus} to
conclude that the imposition of the $SU(1,1)$ nature of $M^c(C_1)$ and
the relation $M^c_{12}(C_2)= \overline{M^c_{21}}(C_2)$ 
is sufficient to assure the $SU(1,1)$ nature of transformations $M^c(C_2)$. As
a consequence $E= M^c_0 M(t)\in SU(1,1)$ and being already $M(t)\in SU(1,1)$ we
have also $M^c_0\in SU(1,1)$. 

\bigskip

\section{Determination of $\beta_1$ and $c_{kj}$}

We come now to the determination of the parameters $\beta_1$ and $c_{jk}$.
The perturbed $Y$ is
\begin{equation}
Y^c=
\begin{pmatrix}
y_1+\delta y_1\\
y_2+\delta y_2
\end{pmatrix}
=(1+F)Y=(1+F)
\begin{pmatrix}
y_1\\
y_2
\end{pmatrix}
\end{equation}
with
\begin{equation}
F=
\begin{pmatrix}
\frac{I_{12}}{w_{12}}+c_{11}&-\frac{I_{11}}{w_{12}}+c_{12}\\
\frac{I_{22}}{w_{12}}+c_{21}& -\frac{I_{12}}{w_{12}}-c_{11}
\end{pmatrix}.
\end{equation}
We denote with $\tilde Y$  and $\tilde Y^c$ the solutions computed 
at the same point on the torus but reached
through a cycle
\begin{equation}
\tilde Y^c=(1+\tilde F)\tilde Y,
\end{equation}
\begin{equation}
\tilde F=
\begin{pmatrix}
\frac{\tilde I_{12}}{w_{12}}+c_{11}&-\frac{\tilde I_{11}}{w_{12}}+c_{12}\\
\frac{\tilde I_{22}}{w_{12}}+c_{21}& -\frac{\tilde I_{12}}{w_{12}}-c_{11}
\end{pmatrix}.
\end{equation}
We have for the unperturbed problem
\begin{equation}
\tilde Y = M Y,~~~~~~~~~~~~~M\in SU(1,1)
\end{equation}
and for the perturbed one
\begin{eqnarray}
\tilde Y^c &=& (1+\tilde F)M Y = (1+\tilde F)M(1-F)Y^c=(M+\tilde F M -M F)Y^c =
\nonumber\\
&=&(I+\tilde F -M FM^{-1})M Y^c= M^c Y^c
\end{eqnarray}
with
\begin{equation}\nonumber
M^c=(I+\tilde F -M FM^{-1})M =M+\delta M.
\end{equation}
We have
\begin{equation}
\tilde F'= \frac{q}{w_{12}}
\begin{pmatrix}
\tilde y_1\tilde y_2&-\tilde y_1 \tilde y_1\\
\tilde y_2 \tilde y_2 & -\tilde y_2\tilde y_1
\end{pmatrix}=
MF'M^{-1}
\end{equation}
and thus
\begin{equation}
\tilde F'-M F'M^{-1} =0
\end{equation}
giving the constancy of $M^c$. 

\bigskip

Writing $M^c=M+\delta M$, for the cycle $C_1$ the relation
$\delta M_{12}(C_1)=\delta \overline {M}_{21}(C_1)$ gives
\begin{eqnarray}\label{rec111}
-2~w_{12}~(c_{11} +\bar c_{11})&=&
I_{12}(C_1)+\overline{I}_{12}(C_1) 
+\tilde I_{12}(C_1)+ \overline{\tilde I}_{12}
(C_1)\nonumber\\
&+&\frac{M_{11}(C_1)}{M_{12}(C_1)}(-w_{12}~(c_{12} -\bar c_{21})
+ I_{11}(C_1) + \overline{I}_{22}(C_1))\nonumber\\
&+&\frac{\overline{M}_{11}(C_1)}{M_{12}(C_1)}\left(w_{12}~
(c_{12} -\bar c_{21}) 
-\tilde I_{11}(C_1)-\overline{\tilde I}_{22}(C_1)\right).
\end{eqnarray}

The $I_{jk}(C_r)$ are of the form
\begin{equation}
I_{jk}(C_r) = \varepsilon A_{jk}(C_r)+\beta_1 B_{jk}(C_r)
\end{equation}
\begin{equation}
\tilde I_{jk}(C_r) = \varepsilon \tilde A_{jk}(C_r)+\beta_1 \tilde B_{jk}(C_r).
\end{equation}
We see that only the combinations 
\begin{equation}
c_{11}+\bar c_{11}\equiv r={\rm real}~~~~{\rm  and}~~~~
c_{12}-\bar c_{21}\equiv s 
\end{equation}
appear in such relation. This, as discussed in section 
\ref{thetorus}, is due to the remnant
$SU(1,1)$ invariance of the monodromy conditions.
Thus the previous is a system of two linear non homogeneous equation 
in $r$, ${\rm Re}~s$, ${\rm Im}~s$ and ${\rm Re}~\beta$, ${\rm Im}~\beta$.

The condition $|M^c_{11}(C_1)|=|M^c_{22}(C_1)|$ gives
\begin{equation}
{\rm Re}(\delta M_{11}(C_1)\overline{M}_{11}(C_1))=
{\rm Re} (\delta M_{22}(C_1)\overline{M}_{22}(C_1))
\end{equation}
explicitely
\begin{eqnarray}\label{modulus1}
& &\overline{M}_{11}(C_1) M_{11}(C_1)
(-I_{12}(C_1)-\overline{I}_{12}(C_1)
+\tilde I_{12}(C_1)+\overline{\tilde I}_{12}(C_1)
)\nonumber\\
&+& {\rm Re}\bigg[\overline{M}_{11}(C_1)M_{12}(C_1)
\big(-I_{22}(C_1) -\overline{I}_{11}(C_1)+w_{12}~\bar s\big)\nonumber\\
&+& \overline{M}_{11}(C_1)\overline{M}_{12}(C_1)
\big(-\tilde I_{11}(C_1)-\overline{\tilde I}_{22}(C_1) +w_{12}~s\big) 
\bigg]=0
\end{eqnarray}
which is one inhomogeneous linear equation.

To this we must add the relation $\delta M_{12}(C_2)=
\delta \overline{M}_{21}(C_2)$ which reads
\begin{eqnarray}\label{rec112}
-2~w_{12}~r&=&
I_{12}(C_2)+\overline{I}_{12}(C_2) 
+\tilde I_{12}(C_2)+ \overline{\tilde I}_{12}
(C_2)\nonumber\\
&+&\frac{M_{11}(C_2)}{M_{12}(C_2)}(-w_{12}~s 
+ I_{11}(C_2) + \overline{I}_{22}(C_2))\nonumber\\
&+&\frac{\overline{M}_{11}(C_2)}{M_{12}(C_2)}\left(w_{12}~s 
-\tilde I_{11}(C_2)-\overline{\tilde I}_{22}(C_2)\right)
\end{eqnarray}
which is an other system of two linear inhomogeneous equations.

The equality of the
real parts of (\ref{rec111}) and (\ref{rec112}) and the
vanishing of the imaginary part of (\ref{rec111}) and
(\ref{rec112}) and (\ref{modulus1}) provide a linear inhomogeneous
system of four equations in the unknown ${\rm Re}~\beta$, ${\rm Im}~\beta$, 
${\rm Re}~s$, ${\rm Im}~s$. The outcome substituted into 
(\ref{rec111}) provides $r$.

\bigskip

\section{The Green function on the one-point background}\label{greenfunction}

Taking the derivative wrt $\varepsilon$ of the equation 
\begin{equation}
-\partial_z\partial_{\bar z}\phi +e^\phi = 2\pi\eta\delta^2(z)
+2\pi\varepsilon\delta^2(z-z_t)
\end{equation} 
satisfied by the
$\phi$ we computed in the previous section we obtain the exact
Green function on the one-point background. In fact we have
\begin{equation}
(-\partial_z\partial_{\bar z}+e^\phi) \frac{1}{2\pi}\frac{d\phi}{d\varepsilon} 
= \delta^2(z-z_t).
\end{equation} 
$\displaystyle{\frac{d\phi}{d\varepsilon}}$ 
can be simply computed by taking the derivative of $\varphi$
obtained in the previous section from $y^c_1$ and $y^c_2$, as the logarithm 
of the Jacobian appearing in the
transition from $\phi$ to $\varphi$ does not depend on $\varepsilon$. We have
adopting the choice $|y_1|^2-|y_2|^2>0$ (we recall that the previous difference
never can vanish) 
\begin{equation}
-\frac{1}{2}\frac{d \varphi}{d\varepsilon}
 e^{-\frac{\varphi}{2}} 
=(\bar y_1 \frac{d \delta y_1}{d\varepsilon}+ c.c.)-
(\bar y_2 \frac{d \delta y_2}{d\varepsilon}+ c.c.)
\end{equation} 
\begin{eqnarray}
G(z,z_t) =\frac{1}{2\pi}\frac{d\varphi}{d\varepsilon}&=&
-\frac{1}{\pi w_{12}(\overline {y_1} y_1-\overline {y_2} y_2)}
\left[(\overline {y_1} y_1+\overline {y_2} y_2)\left(\dot I_{12}
+\overline{\dot I}_{12}+w_{12}~\dot r\right)\right.\nonumber\\
&+&\left.\left(\overline {y_1} y_2(-\dot I_{11}
-\overline{{\dot I}}_{22}+w_{12}~\dot s) + c.c.\right)\right]
\end{eqnarray} 
where
\begin{equation}
\dot I_{jk}=\frac{I_{jk}}{\varepsilon},~~~~\dot r =\frac{r}{\varepsilon}~,
~~~~\dot s =\frac{s}{\varepsilon}
\end{equation} 
and $G$ satisfies
\begin{equation}
(-\partial_z\partial_{\bar z} +e^\phi)G(z,z_t) = \delta^2(z-z_t).
\end{equation}
 
\section{Conclusions}\label{conclusions}

In this paper we have extended the treatment of a previous paper
\cite{menottitorus} to the general case in which the distribution of charges
is not symmetric under reflections with respect to the origin. In this case
one has to apply the concept of differential equation on a Riemann surface and
in particular the concept of Fuchsian differential equation on a Riemann
surface. As in \cite{menottitorus} new accessory parameters appear which have
to be fixed by imposing the monodromic behavior of the solution. This is a
more difficult task than in the symmetric case where sphere topology
techniques could be applied. We gave a general group theoretic argument to
prove that the Heun parameter and the weight factor are sufficient to satisfy
all the monodromies. To give a concrete illustration of the procedure, in
Appendix A we work out explicitely a perturbative case. Then in section
\ref{additionweak} we deal with the problem of the addition of a weak source
to the solution of the one point function. The procedure is completely general
even if we know the explicit form of the solution only in the cases of the
square and of the rhombus of opening angle $2\pi/6$ which are reported in
Appendix B. This is due to the fact that in these cases for symmetry reasons
the Heun parameter $\beta$ vanishes. As a byproduct we also obtain the exact
expression in terms of quadratures of the Green function on the background of
one source of arbitrary strength. The results can be used to develop the
quantum semiclassical expansion as done for the pseudosphere, sphere and disk
topologies in \cite{MT}.

\section*{Appendix A}

In this appendix we summarize the general perturbative treatment of the one
point function on the torus when the source is not symmetrical with respect
to the kinematical singularities $e_k$ and which requires explicitely the two
sheet representation of the torus. The general discussion in the non
perturbative case was given in section \ref{thetorus}; here we give the
explicit calculation for the perturbative case. 
The differential equation in the 
variable $u$ is
given by
\begin{equation}
\frac{d^2y}{du^2} +Q y =0
\end{equation}
with $Q = Q_0 +q$ and
\begin{eqnarray}\label{Q0}
Q_0(u)&=&\frac{3}{16}\left(\frac{1}{(u-e_1)^2}+ \frac{1}{(u-e_2)^2}+
\frac{1}{(u-e_3)^2}+\frac{2 e_1}{(e_1-e_2)(e_3-e_1)(u-e_1)}\right.\nonumber\\
&+& \left.\frac{2 e_2}{(e_2-e_3)(e_1-e_2)(u-e_2)} 
+\frac{2 e_3}{(e_3-e_1)(e_2-e_3)(u-e_3)}\right) \nonumber\\
&=&-\{z,u\} = \{u,z\}\left(\frac{du}{dz}\right)^{-2}.
\end{eqnarray}
$Q_0$ is simply minus the Schwarzian derivative $\{z,u\}$ for the transition 
from the covering
variable $z$ which describes the torus by the plane modulo the discrete
translation group, to the variable $u=\wp(z)$. Thus the $e_j$ are
``kinematical'' singularities.
 
The term $q$  is a meromorphic function on the cubic curve 
(\ref{cubic})
describing the weak source at the point $(t,w_t)$  
\begin{eqnarray}\label{smallqpert}
q(u,w) &=&\frac{\varepsilon}{w_t}
\left[\frac{(w+w_t)^2}{4(u-t)^2}
-u-2t\right]\frac{1}{w} 
+\frac{\beta_t(w+w_t)}{2(u-t)w}\nonumber\\
&+&\frac{\beta_1}{2(u-e_1)}+\frac{\beta_2}{2(u-e_2)}+\frac{\beta_3}{2(u-e_3)}.
\end{eqnarray}
As we shall see later we have to allow in (\ref{smallqpert}) 
in order to satisfy the monodromy
conditions, for the new accessory parameter $\beta_t$ in addition to the
variations $\beta_j$ of the unperturbed accessory parameter appearing in
(\ref{Q0}).

The Fuchs relations which leave the $u=\infty$ behavior regular with no
source at $z=0$ are
\begin{equation}\label{fuchspert}
0=\beta_t+\beta_1+\beta_2+\beta_3,~~~~
0=\varepsilon +t~\beta_t+e_1~\beta_1+e_2~\beta_2+e_3~\beta_3.
\end{equation}

Using $v$, with $v^2=1/u$ as covering variable at $u=\infty$, the asymptotic
behavior of $q$ is given by
\begin{eqnarray}\label{vasymptotic}
q &=& \frac{\beta_1+\beta_2+\beta_3+\beta_t}{2}v^2+
\frac{\beta_1 e_1+\beta_2 e_2 +\beta_3 e_3 +\beta_t t +
\varepsilon}{2}v^4+\nonumber\\
&+&\frac{2\beta_t w_t+\varepsilon(12 t^2-g_2)/w_t}{8} v^5+\nonumber\\
&+&\frac{\beta_1 e_1^2+\beta_2 e_2^2 +\beta_3 e_3^3 +\beta_t t^2 +2
~\varepsilon~ t}{2}v^6+O(v^7).
\end{eqnarray}

First we examine the monodromy at $(t,w_t)$. 
Two independent solutions of the
unperturbed equation $y''+Q_0 y=0$ are
\begin{equation}
y_1 = \sqrt{\frac{w}{2}},~~~~y_2 = \sqrt{\frac{w}{2}}~ (Z-\omega_3),~~~~{\rm
with}~~~~ Z = z-z_t .
\end{equation}
Their Wronskian is $w_{12}=1/2$. 
The perturbed $y_1$ solution is given by
\begin{equation}
y_1+\delta y_1= y_1+y_1 \frac{I_{12}}{w_{12}} -y_2\frac{I_{11}}{w_{12}}
\end{equation}
with
\begin{eqnarray}
I_{11}&=&\int_{u_0}^u q(u)  y_1(u)y_1(u)du=\\
&&\frac{1}{2}\int_{u_0}^u \left(\frac{\varepsilon}{w_t}
\left[\frac{(w+w_t)^2}{4(u-t)^2}-u-2t\right]\frac{1}{w}+
\frac{\beta_t(w+w_t)}{2(u-t)w}+
\sum_k\frac{\beta_k}{2(u-e_k)}\right) w du \nonumber~.
\end{eqnarray}
A necessary condition for having monodromy at $(t,w_t)$ is that the change of
$I_{11}$ for a tour around $t$ vanishes i.e.  
\begin{equation}\label{betatfix}
0=\delta I_{11} = \frac{2\pi i}{2}\left(w_t\beta_t+
\varepsilon\frac{12 t^2-g_2}{2w_t}\right).
\end{equation}
We notice that such a condition on $\beta_t$ gives also the monodromy at 
$v=0$ as
from (\ref{vasymptotic}) we have that the first order pole at $v=0$ of
$q(u,w)w ~du/dv$
vanishes. This is an outcome of the fact that the sum of the residues of a
meromorphic function on a
closed manifold vanishes and due to the factor $w$ there are no residues
at the points $(e_j,0)$. After fixing $\beta_t$ to satisfy (\ref{betatfix}), 
due to the two Fuchs
conditions (\ref{fuchspert}) we have still one accessory parameter free.

In absence of the first order poles $I_{11}$, due to a general theorem
\cite{batemanII} becomes
with $Z=z-z_t$
\begin{equation}
I_{11} = \frac{\varepsilon}{2}\int_{\omega_3}^Z(\wp(Z)+c_1)dZ=
\frac{1}{2}\varepsilon
(-\zeta(Z)+\zeta(\omega_3))+\varepsilon\frac{c_1}{2}(Z-\omega_3) 
\end{equation}
where we choose from now on $u_0 =\wp(z_t+\omega_3)$ and $c_1$ replaces the 
free accessory parameter.
Similarly
\begin{eqnarray}
I_{12} &=& \frac{\varepsilon}{2}\int_{\omega_3}^Z
(\wp(Z)+c_1)(Z-\omega_3) dz=\\
&=&-\frac{\varepsilon}{2}\zeta(Z)(Z-\omega_3)
+\varepsilon
\frac{\zeta(\omega_1)}{4\omega_1}(Z^2-\omega_3^2)+\frac{1}{2}
\varepsilon {\cal L}
+\varepsilon \frac{c_1}{4}(Z-\omega_3)^2\nonumber
\end{eqnarray}
where
\begin{equation}
{\cal L}=\log\frac{\theta_1(\frac{Z}{2\omega_1}|\tau)}
{\theta_1(\frac{\omega_3}{2\omega_1}|\tau)}.
\end{equation}
Then
\begin{eqnarray}
& &y^c_1 =y_1+\delta y_1 = \\
& &\sqrt{\frac{w}{2}}\left(1
+\frac{\varepsilon}{w_{12}}
\left\{-\frac{1}{2}
\zeta(\omega_3)(Z-\omega_3)-\frac{c_1}{4}
(Z-\omega_3)^2+\frac{\zeta(\omega_1)}{4\omega_1}
(Z^2-\omega_3^2)+
\frac{1}{2}{\cal L}\right\}\right)\nonumber.
\end{eqnarray}
The remaining condition for the monodromic behavior at $(t,w_t)$ is the
reality of $\varepsilon$ which corresponds to the ellipticity of the
singularity. The perturbed 
$e^{-\frac{\varphi}{2}}$ is \cite{menottitorus}
\begin{equation}
e^{-\frac{\varphi}{2}} =\frac{1}{\sqrt{2}|w_{12}|\kappa|^2}
\left[\overline{y^c_1}y^c_1-
|\kappa|^4 \overline{y^c_2} y^c_2\right]
\end{equation}
where $|\kappa|^4 =O(\varepsilon)$ and thus $y^c_2$ is a general solution 
of the unperturbed equation which we write as 
\begin{equation}
y^c_2 = \sqrt{\frac{w}{2}} (Z-\omega_0)~.
\end{equation}
$\omega_0$ is going to be chosen along with $c_1$ and $|\kappa|$ to satisfy 
the periodicity conditions.

We are left with the imposition of the monodromic behavior along the two 
fundamental non
contractible cycles. This can be done both in the $(u,w)$ or $z$ coordinates.
To order $O(\varepsilon)$ for the structure
$
|y^c_1|^2 - |\kappa|^4 |y^c_2|^2
$
to be monodromic under the transformation
\begin{equation}
y^c_1\rightarrow y^c_1(1+\varepsilon a) + \varepsilon b y^c_2~,~~~~
y^c_2\rightarrow y^c_2 + f y^c_1
\end{equation}
we have the necessary and sufficient conditions
\begin{equation}
\varepsilon (a+\bar a) =|\kappa|^4 \bar f f,~~~~
\varepsilon b=|\kappa|^4 \bar f~.
\end{equation}
Using \cite{batemanII}
\begin{equation}
\theta_1(v-1|\tau) = -\theta_1(v|\tau),~~~~
\theta_1(v-\tau|\tau) = -\theta_1(v|\tau)e^{i\pi(2v-\tau)}
\end{equation}
for the cycle $C_1$, $z\rightarrow z-2\omega_1$ we have for small
$\varepsilon$
\begin{equation}
a ~w_{12}= \omega_1\zeta(\omega_3)+c_1 \omega_1(-\omega_1+
\omega_0-\omega_3)
+\zeta(\omega_1)(\omega_1 -\omega_0) +\frac{1}{2}\log(-1),\nonumber
\end{equation}
\begin{equation}
b~w_{12}= c_1\omega_1 -\zeta(\omega_1),~~~~ f= -2\omega_1
\end{equation}
and for the cycle $C_2$, $z\rightarrow z-2\omega_2$
\begin{equation}
a ~w_{12}=
c_1(-\omega_2^2-\omega_2\omega_3+\omega_0\omega_2)+\omega_2\zeta(\omega_3)+
\zeta(\omega_1)(\frac{\omega_2^2}{\omega_1}-\frac{\omega_2\omega_0}{\omega_1})
+\frac{1}{2}\log(-1)+i\frac{\pi}{2}\frac{\omega_0}{\omega_1}
-i\frac{\pi\tau}{2},\nonumber  
\end{equation}
\begin{equation}
b~w_{12}= c_1\omega_2-\zeta(\omega_1)\frac{\omega_2}{\omega_1} 
+i\frac{\pi}{2\omega_1
},~~~~f = -2\omega_2.
\end{equation}
The equation
\begin{equation}\label{ratios}
\left(\frac{b}{\bar f}\right)_{C_1}=\left(\frac{b}{\bar
f}\right)_{C_2}
\end{equation}
fixes $c_1$ and gives using the Legendre relation with 
${\rm Im}(\omega_2/\omega_1)>0$
\cite{batemanII}
\begin{equation}
c_1 = \frac{\zeta(\omega_1)\bar\omega_2-\zeta(\omega_2)\bar\omega_1}
{\omega_1\bar\omega_2-\bar\omega_1\omega_2}~.
\end{equation}
Then from
\begin{equation}
|\kappa|^4 =\frac{\varepsilon b}{\bar f}
\end{equation}
we have
\begin{equation}
|\kappa|^4
=\frac{\varepsilon}{w_{12}}\frac{\zeta(\omega_2)\omega_1
-\zeta(\omega_1)\omega_2}
{2(\omega_1\bar\omega_2-\bar\omega_1\omega_2)}=
-\frac{i\pi\varepsilon}{4w_{12}(\omega_1\bar\omega_2-\bar \omega_1\omega_2)}
>0~.
\end{equation}
Imposition of the remaining relations involving the $a$'s 
give $\omega_0=\omega_3$ and the final
result is
\begin{eqnarray}
&&e^{-\frac{\phi}{2}}=\frac{1}{2\sqrt{2}w_{12}|\kappa|^2}\bigg(1+
\frac{\varepsilon}{2w_{12}}\bigg[\log\left|\frac{\theta_1
(\frac{Z}{2\omega_1}|\tau)} 
{\theta_1(\frac{\omega_3}{2\omega_1}|\tau)}\right|^2\nonumber\\
&-&\frac{i\pi}{4(\omega_1\bar\omega_2-\bar\omega_1\omega_2)}
(\frac{\bar\omega_1}{\omega_1}Z^2+
\frac{\omega_1}{\bar\omega_1}\bar Z^2-2Z\bar Z)
+\frac{i\pi}{4}(\frac{\bar\omega_2}{\bar\omega_1}-
\frac{\omega_2}{\omega_1})\bigg]\bigg).
\end{eqnarray} 
The relation between the
$\varepsilon$ of the previous paper \cite{menottitorus} (call it
$\varepsilon_p$) and the present $\varepsilon$ is $\displaystyle{
\frac{\varepsilon_p}{4} = \varepsilon}$.

\section*{Appendix B}

For the reader's convenience we report in this appendix the exact solution for
the square given in \cite{menottitorus} in terms of hypergeometric functions. 
We add
also the explicit solution in terms of hypergeometric functions of an other
soluble case i.e. when the torus is a rhombus with an opening angle of $2
\pi/6$. 

1) The square.

With $e_3=0, e_1=-e_2=1$ and $\beta=0$ the term $Q$ of eq.(\ref{Qterm})
becomes
\begin{equation}
Q(u) = \frac{1-\lambda^2}{16(u^2-1)}+\frac{3}{16}\frac{(1+u^2)^2}
{u^2(1-u^2)^2}.
\end{equation}

Two independent solutions canonical at $e_3=0$ are
\begin{eqnarray}\label{y1y2hyper}
y_1&
=&\kappa^{-1}u^{\frac{1}{4}}(1-u^2)^{\frac{1}{4}}
F(\frac{1-\lambda}{8},\frac{1+\lambda}{8};\frac{3}{4};u^2)\nonumber\\
y_2&=&\kappa ~u^{\frac{3}{4}}(1-u^2)^{\frac{1}{4}}
F(\frac{3-\lambda}{8},\frac{3+\lambda}{8};\frac{5}{4};u^2)~.
\end{eqnarray}
and the parameter $|\kappa|$ is 
\begin{equation}\label{kappa4square}
|\kappa|^4 =\left(\frac{\Gamma(\frac{3}{4})}{\Gamma(\frac{5}{4})}\right)^2 
~\frac{\Gamma(\frac{3-\lambda}{8})
\Gamma(\frac{3+\lambda}{8}) \Gamma(\frac{7-\lambda}{8})
\Gamma(\frac{7+\lambda}{8})}
{\Gamma(\frac{1-\lambda}{8})
\Gamma(\frac{1+\lambda}{8}) \Gamma(\frac{5-\lambda}{8})
\Gamma(\frac{5+\lambda}{8})}\equiv
8~\frac{\Gamma^2(\frac{3}{4})\Gamma(\frac{3-\lambda}{4})   
\Gamma(\frac{3+\lambda}{4})}
{\Gamma^2(\frac{1}{4})\Gamma(\frac{1-\lambda}{4})
\Gamma(\frac{1+\lambda}{4})}
\end{equation}
so that the conformal factor is given by
\begin{equation}
e^{-\frac{\phi(z)}{2}}=\frac{1}{\sqrt{2}|\kappa|^2}
\left[\left|F(\frac{1-\lambda}{8},\frac{1+\lambda}{8};\frac{3}{4};u^2(z))
\right|^2- |\kappa|^4 |u(z)|
\left|F(\frac{3-\lambda}{8},\frac{3+\lambda}{8};\frac{5}{4};u^2(z))\right|^{2}
\right]
\end{equation}
with $u(z)=\wp(z)$ and the half periods
corresponding to $e_1 = -e_2=1,~e_3=0$ are
\begin{equation}
\omega_1 =-i \omega_2 = \sqrt{\pi}\frac{\Gamma(\frac{5}{4})}
{\Gamma(\frac{3}{4})}= 1.31103..
\end{equation}

\bigskip

2) The rhombus with opening angle $2\pi/6$.

Also in this case for symmetry reasons combined with Picard's uniqueness 
theorem
we have that $\beta=0$ in the $Q$ of eq.(\ref{Qterm}). The values of 
the $e_k$ are
given by $e_1=1$, $e_2 = \exp(2 i \pi/3)$, $e_3 = \exp(4 i\pi/3)$. 
The term $Q$ of eq.(\ref{Qterm})
becomes
\begin{equation}
Q(u)= \frac{1-\lambda^2}{16}\frac{u}{u^3-1}+\frac{3}{16}
\left(\frac{3u^4+6u}{(u^3-1)}-\frac{2u}{u^3-1}\right)~.
\end{equation}
Going over to the variable $x=u^3$ we obtain the differential equation
\begin{equation}
9x\frac{d^2y}{dx^2}+6\frac{dy}{dx}+\left[\frac{1-\lambda^2}{16(x-1)}
+\frac{3}{16}
\left(\frac{3x+6}{(x-1)^2}-\frac{2}{x-1}\right)\right] y=0
\end{equation}
which can be solved in terms of hypergeometric functions. Two independent
solutions canonical at $u=0$ are given by
\begin{eqnarray}
y_1(u)&=& \kappa^{-1}(1-u^3)^\frac{1}{4}
F(\frac{1-\lambda}{12},\frac{1+\lambda}{12};\frac{2}{3};u^3)\nonumber\\
y_2(u)&=& \kappa u(1-u^3)^\frac{1}{4}
F(\frac{5-\lambda}{12},\frac{5+\lambda}{12};\frac{4}{3};u^3).
\end{eqnarray}
Monodromicity conditions impose 
\begin{equation}\label{kappa4rhombus}
|\kappa|^4 = \left(\frac{\Gamma(\frac{2}{3})}{\Gamma(\frac{4}{3})}\right)^2
\frac{\Gamma(\frac{5-\lambda}{12})\Gamma(\frac{5+\lambda}{12})
\Gamma(\frac{11-\lambda}{12})\Gamma(\frac{11+\lambda}{12})}
{\Gamma(\frac{7-\lambda}{12})\Gamma(\frac{7+\lambda}{12})
\Gamma(\frac{1-\lambda}{12})\Gamma(\frac{1+\lambda}{12})}\equiv
9~\frac{\pi \Gamma(\frac{5-\lambda}{6})\Gamma(\frac{5+\lambda}{6})}
{\Gamma^2(\frac{1}{6})\Gamma(\frac{1-\lambda}{6})\Gamma(\frac{1+\lambda}{6})}
\end{equation}
from which we obtain the conformal factor
\begin{equation}
e^{-\frac{\phi(z)}{2}}=\frac{1}{\sqrt{2}|\kappa|^2}
\left[\left|F(\frac{1-\lambda}{12},\frac{1+\lambda}{12};\frac{2}{3};u^3(z))
\right|^2- |\kappa|^4 |u(z)^2|
\left|F(\frac{5-\lambda}{12},\frac{5+\lambda}{12};
\frac{4}{3};u^3(z))\right|^{2}
\right]\nonumber
\end{equation}
where $|\kappa|^4$ is given by eq.(\ref{kappa4rhombus}) and the half periods 
corresponding to
$e_1=1$, $e_2=\exp(2 i \pi/3)$, $e_3=\exp(4 i \pi/3)$ are
\begin{equation} 
\omega_1 = \omega_2 \exp(-i\pi/3)= 
\sqrt{\pi}\frac{\Gamma(\frac{7}{6})}{\Gamma(\frac{2}{3})}= 1.21433..
\end{equation} 

\eject

\vfill


\end{document}